\documentclass[conference]{IEEEtran}
\ifCLASSINFOpdf
\else
\fi
\usepackage{times}
\usepackage{cite}
\usepackage[cmex10]{amsmath}
\usepackage{array}
\usepackage{mdwmath}
\usepackage{mdwtab}
\usepackage{eqparbox}
\usepackage{stfloats}
\usepackage{graphicx}
\usepackage{float}
\usepackage{url}
\usepackage{color}
\usepackage{setspace}
\usepackage{tikz}
\usetikzlibrary{patterns}
\usepackage{pgfplots}
\usepackage{pgfplotstable}
\pgfdeclarelayer{background}
\usetikzlibrary{spy,fit}
\pgfsetlayers{background,main}
\usetikzlibrary{shapes,decorations,calc}
\usetikzlibrary{matrix,decorations.markings}
\usetikzlibrary{arrows}

\tikzstyle{every picture}+=[task/.style={draw,circle}, fork/.style={draw}]
\pgfplotscreateplotcyclelist{black white}{%
  densely dotted,every mark/.append
    style={solid,fill=gray},mark=diamond*\\bmod@
    every mark/.append style={fill=gray},mark=*\\bmod@
    densely dashed,every mark/.append
    style={solid,fill=gray},mark=+\\bmod@
    densely dotted,every mark/.append
    style={solid,fill=gray},mark=star\\bmod@
}
\pgfplotsset{cycle list name=black white}

\usepackage[ruled,vlined]{algorithm2e}

\DeclareMathSizes{10.5}{18.5}{10}{9}
\everymath{\displaystyle}

\SetAlFnt{\small}
\SetKwFunction{MPIB}{MPI\_Bcast}
\SetKwFunction{CPBA}{Copy\_Block\_A}
\SetKwFunction{CPBB}{Copy\_Block\_B}
\SetKwFunction{Prow}{Pivot\_row}
\SetKwFunction{Pcol}{Pivot\_col}
\SetKwFunction{MPIC}{MPI\_Comm}
\SetKwFunction{DGemm}{DGemm}
\SetKwFunction{CShift}{CIRCULAR\_SHIFT}

\newcommand{\sq}{\hbox{\rlap{$\sqcap$}$\sqcup$}}
\newcommand{\qed}{\hspace*{\fill}\sq}

\begin{document}
%
\title{Hierarchical Parallel Matrix Multiplication on Large-Scale Distributed Memory Platforms}

\author{\IEEEauthorblockN{Jean-No\"el Quintin}
\IEEEauthorblockA{Extrem Computing R\&D\\
Bull\\
France\\
jean-noel.quintin@bull.net}
\and
\IEEEauthorblockN{Khalid Hasanov}
\IEEEauthorblockA{University College Dublin\\
Dublin 4, Belfield\\
Ireland\\
khalid.hasanov@ucdconnect.ie}
\and
\IEEEauthorblockN{Alexey Lastovetsky}
\IEEEauthorblockA{University College Dublin\\
Dublin 4, Belfield\\
Ireland\\
Alexey.Lastovetsky@ucd.ie}}


%


\maketitle

\begin{abstract}
Matrix multiplication is a very important computation kernel both in its own right as a building
block of many scientific applications and as a popular representative for other
scientific applications.

Cannon's algorithm which dates back to 1969 was the first efficient algorithm for
parallel matrix multiplication providing theoretically optimal communication cost.
However this algorithm requires a square number of processors.
In the mid-1990s, the SUMMA algorithm was introduced. SUMMA overcomes the shortcomings of Cannon's algorithm
as it can be used on a non-square number of processors as well. Since then the number of processors
in HPC platforms has increased by two orders of magnitude making the contribution of communication in
the overall execution time more significant. Therefore, the state of the art parallel matrix multiplication algorithms
should be revisited to reduce the communication cost further.

This paper introduces a new parallel matrix multiplication algorithm, Hierarchical SUMMA (HSUMMA),
which is a redesign of SUMMA. Our algorithm reduces the communication cost of SUMMA by
introducing a two-level virtual hierarchy into the two-dimensional arrangement of processors.
Experiments on an IBM BlueGene/P demonstrate the reduction of communication cost up to $2.08$
times on $2048$ cores and up to $5.89$ times on $16384$ cores.

\end{abstract}
%
\IEEEpeerreviewmaketitle

\section{Introduction}
Matrix multiplication is a very important computation kernel both in its own right as a building
block of many scientific applications and as a popular representative for other scientific applications.

Cannon's algorithm\cite{cannon:mmm} which was introduced in 1967 was the first efficient algorithm for parallel matrix multiplication providing theoretically optimal 
communication cost. However this algorithm requires a square number of processors which makes it impossible to be used in a general purpose library.
Later introduced Fox's algorithm\cite{fox:mmm} has the same restriction.

The 3D algorithm\cite{3d:mmm} which dates back to the 1990s organizes the $p$ processors as $p^{\frac{1}{3}}{\times}p^{\frac{1}{3}}{\times}p^{\frac{1}{3}}$ 3D mesh
and achieves a factor of $p^{\frac{1}{6}}$ less communication cost than 2D parallel matrix multiplication algorithms. However, in order to get this improvement 
the 3D algorithm requires $p^{\frac{1}{3}}$ extra copies of the matrices. That means that on one million cores the $3D$ algorithm will require $100$ extra copies of the matrices
which would be a significant problem on some platforms. Therefore, the 3D algorithm is only practical for relatively small matrices. 

One implementation of Fox's algorithm on a general $P{\times}Q$ processor grid is PUMMA\cite{ChoiWD94:0} which was designed for block-cyclic distributed matrices. PUMMA
consists of $Q-1$ shifts for matrix $A$, $LCM(P,Q)$ broadcasts for matrix $B$ and the number of local multiplications is $LCM(P,Q)$. Here $LCM(P,Q)$ is the least
common multiple of $P$ and $Q$. The main shortcomings of PUMMA come from the fact that it always tries to use the largest possible matrices for both computation
and communication. In this case, large memory space is required to store them temporarily, the effect of the block size is marginal and the most important it is
difficult to overlap computation with communication.

In the mid-1990s SUMMA\cite{summa:mmm} was introduced. Like PUMMA, SUMMA was designed for a general $P{\times}Q$ processor grid. Unlike PUMMA it does not require the largest
possible matrices for computation and communication and therefore allows to pipeline them. In addition, SUMMA was implemented in practice in ScaLAPACK\cite{Dongarra:1997:SUG:265932}:
the most popular parallel numerical linear algebra package.

Recently introduced 2.5D algorithm\cite{Solomonik:2011:CPM:2033408.2033420} generalizes the 3D algorithm
by parametrizing the extent of the third dimension of the processor arrangement: $\frac{p}{c}^{\frac{1}{2}}{\times}\frac{p}{c}^{\frac{1}{2}}{\times}c$,
$c \in [1, p^{\frac{1}{3}}]$. However, the 2.5D algorithm is efficient only if there is free amount of extra memory to store $c$ copies of the matrices.
On the other hand, it is expected that exascale systems will have a dramatically shrinking memory space per core\cite{exascale:report}. 
Therefore, the 2.5D algorithm can not be scalable on the future exascale systems.

Matrix multiplication is a problem known to be very compute intensive. On the other hand, as HPC moves towards exascale, 
the cost of matrix multiplication will be dominated by communication cost. Therefore, the state of the art parallel matrix multiplication algorithms
should be revisited to reduce the communication cost further. 

The contributions of this paper are as follows:
\begin{itemize}
\item
We introduce a new design to parallel matrix multiplication algorithm by introducing a two-level virtual hierarchy into the
two-dimensional arrangement of processors. We apply our approach to SUMMA which is a state of the art algorithm. We call our algorithm hierarchical SUMMA(HSUMMA).
\item
We model the performance of SUMMA and HSUMMA and theoretically prove that HSUMMA reduces the communication cost of SUMMA. 
Then we provide experimental results on a cluster of Grid5000 and BlueGene/P which are reasonable representative and span a good spectrum of
loosely and tightly coupled platforms. We use SUMMA as the only competitor to our algorithm because it is the most general and scalable 
parallel matrix multiplication algorithm, which decreases its per-processor memory footprint with the increase of the number of processors for a given problem size, 
and is used in the most famous parallel numerical linear algebra packages such as ScaLAPACK. 
In addition, because of its practicality SUMMA plays a starting point to implement parallel matrix multiplications on specific platforms. As a matter of fact,
the most used parallel matrix multiplication algorithm for heterogeneous platforms\cite{Beaumont01matrixmultiplication}\cite{alexey:hphc} was based on SUMMA as well.
Therefore, despite it was introduced in the mid-1990s SUMMA is still a state of the art algorithm.
\end{itemize}



\section{Previous Work}
In this section, we detail SUMMA algorithm which is the motivating work for our algorithm. Then, we outline and discuss the existing 
broadcast algorithms which can be used inside SUMMA to improve its communication cost.

\subsection{SUMMA algorithm}
\label{Sec:summa}
SUMMA\cite{summa:mmm} implements the matrix multiplication $C = A\times{B}$ over a two-dimensional $p=s\times{t}$ processor grid.
For simplicity, let us assume that the matrices are square $n\times{n}$ matrices. These matrices are distributed over the processor grid by block-distribution.

\begin{figure}[H]
\begin{center}
\begin{tikzpicture}
\coordinate (org) at (0,0);
\draw[step=0.7] ($(org) + (0,0)$) grid ++(2.8,2.8)    ;
\draw[pattern = north west lines] ($(org) + (1.4,0)$) rectangle ++(0.7,2.8);
\foreach \i in {0,0.7,1.4,2.1}
{
\draw[ <-, double distance=1pt
        ] ($(org) + (2.5,\i+.3)$) -- ++(-0.6,0);
\draw[  ->, double distance=1pt
        ] ($(org) + (1.5,\i+.3)$) -- ++(-0.6,0);
\draw[  <-, double distance=1pt
        ] ($(org) + (0.3,\i+.3)$) .. controls ++(0.5,0.5) and ++(-0.5,0.5) ..
        ($(org) + (1.7,\i+.2)$);
}
\end{tikzpicture}
\begin{tikzpicture}[rotate=180]
\coordinate (org) at (3,0);
\draw[step=0.7,shift={(3,0)}] ($(org) + (0.0,0)$) grid ++(2.8,2.8);
\draw[pattern = north west lines] ($(org) + (0,1.4)$) rectangle ++(2.8,0.7);

\foreach \i in {0,0.7,1.4,2.1}
{
\draw[ <-, double distance=1pt
        ] ($(org) + (\i+.3,0.2)$) ..controls ++(0.5,0.5) and ++(0.3,-0.3) ..
        ++(0.1,1.4);
\draw[ <-, double distance=1pt
        ] ($(org) + (\i+.5,1.0)$) -- ++(0,0.6);
\draw[-> , double distance=1pt
        ] ($(org) + (\i+.5,2.0)$) -- ++(0,0.6);
}
\end{tikzpicture}
\end{center}
\caption{Communication Pattern of SUMMA for the third step with four processors and one block per processor}
\label{comm:SUMMA}
\end{figure}
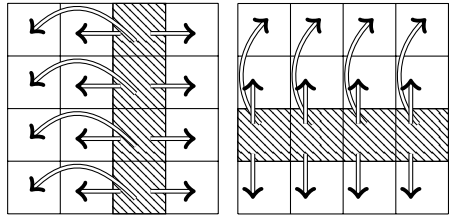

We can see the size of the matrices as $\frac{n}{b}{\times}\frac{n}{b}$ by introducing a block of size $b$. Then each element in $A$, $B$, and $C$
is a square $b{\times}b$ block, and the unit of computation is the updating of one block,
that is, a matrix multiplication of size $b$. For simplicity, we assume that $n$ is a multiple of $b$.
The algorithm can be formulated as follows:
The algorithm consists of $\frac{n}{b}$ steps. At each step
\begin{itemize}
\item
Each processor holding part of the pivot column of the matrix $A$ horizontally broadcasts its part of the pivot column along processor row.
\item
Each processor holding part of the pivot row of the matrix $B$ vertically broadcasts its part of the pivot row along processor column.
\item
Each processor updates each block in its $C$ rectangle with one block from the pivot column and one block from the pivot row,
so that each block $c_{ij}, (i, j)\in(1,...,\frac{n}{b})$ of matrix $C$ will be updated as $c_{ij}=c_{ij}+a_{ik}{\times}{b_{kj}}$.
\item
After $\frac{n}{b}$ steps of the algorithm, each block $c_{ij}$ of matrix $C$ will be equal to $\sum_{k=1}^{\frac{n}{b}}a_{ik}\times{b_{kj}}$
\end{itemize}
Figure~\ref{comm:SUMMA} shows the communication pattern at the third step with one block per processor.

\subsection{MPI Broadcast Algorithms}
Collective communications are fundamental tools to parallelize matrix multiplication algorithms. 
We have already seen that the communication pattern of SUMMA is based on broadcast and an improvement in the broadcast algorithm can 
improve the communication cost of SUMMA as well. Therefore it is worth to outline existing broadcast algorithms.

A lot of research have been done into MPI\cite{Snir:1998:MCR:552013} collective communications and especially into MPI broadcast algorithms
\cite{Thakur03improvingthe}\cite{Barnett94interprocessorcollective}\cite{Thakur05optimizationof}. Early implementations of broadcast algorithms assumed homogeneous
and fully connected networks. They are based on simple binary or binomial trees. On the other hand, some algorithms have been introduced in order to be more effective
for large message sizes and use the benefits of hierarchical networks by using pipelined trees or recursive halving algorithms\cite{Thakur03improvingthe}.
However, neither the broadcast APIs nor numerous broadcast algorithms are application specific, and most of the time improvements come from platform 
parameters and very often they are for specific architectures, such as mesh, hypercube and fat tree\cite{633174}.

In the next section we introduce hierarchical SUMMA algorithm. Our algorithm is neither an improvement of an existing broadcast algorithm nor a new broadcast algorithm. 
HSUMMA is an application specific but platform independent hierarchical matrix multiplication algorithm which reduces communication cost of 
SUMMA and allows better overlapping of communications and computation. Indeed, HSUMMA can use any of the existing optimized broadcast algorithms and still
reduce the communication cost of SUMMA as demonstrated in Section~\ref{Sec:general_model}.

\section{Hierarchical SUMMA}
\label{Sec:hsumma}
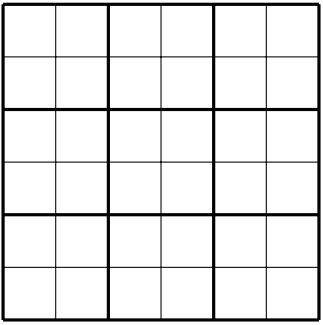
\begin{figure}[H]
\begin{center}
\begin{tikzpicture}[transform shape, scale=0.7]
  \draw (0, 0) grid (6, 6);
  \draw[very thick] (0, 0) grid[step=2] (6, 6);
\end{tikzpicture}
\end{center}
\caption{Hierarchical platform as nine groups, four processors per group}
\label{grid:hierarchical}
\end{figure}

Let us assume we have $p=s\times{t}$ processors distributed over the same two-dimensional virtual processor grid as in SUMMA,
the matrices are square $n\times{n}$ matrices, $b$ is the block size. The distribution of the matrices is the same as in SUMMA.
HSUMMA partitions the virtual $s\times{t}$ processor grid into a higher level $I\times{J}$ arrangement of rectangular groups of processors,
so that inside each group there is a two-dimensional $\frac{s}{I}\times{\frac{t}{J}}$ grid of processors.
Figure~\ref{grid:hierarchical} gives an example of such two-level hierarchical arrangement of processors. In this example a $6{\times}6$ grid of
processors is arranged into two-level $3{\times}3$ grids of groups and $2{\times}2$ grid of processors inside a group.
Let $P_{(x,y)(i,j)}$ denote the processor (i,j) inside the group (x,y).
HSUMMA splits the communication phase of the SUMMA algorithm into two phases and consists of $\frac{n}{b}$ steps.
The algorithm can be summarized as follows:
\begin{itemize}
\item
Horizontal broadcast of the pivot column of the matrix $A$ is performed as follows:
\begin{itemize}
\item
First, each processor $P_{(k,y)(i,j)}, k\in(1,...,I)$ holding part of the pivot column of the matrix $A$ horizontally broadcasts its part of the pivot column
to the processors $P_{(k,z)(i,j)}$, $z{\neq}y, z\in(1,...,I)$ in the other groups.
\item
Now, inside each group $(x,y)$ processor $P_{(x,y)(i,j)}$ has the required part of the pivot column of the matrix $A$ and it further horizontally broadcasts it to the processors $P_{(x,y)(i,c)}$,
$c{\neq}j, c\in(1,...,\frac{s}{I})$ inside the group.
\end{itemize}
\item
Vertical broadcast of the pivot row of the matrix $B$ is performed as follows:
\begin{itemize}
\item
First, each processor $P_{(x,k)(i,j)}, k\in(1,...,I)$ holding part of the pivot row of the matrix $B$ vertically broadcasts its part of the pivot row
to the processors $P_{(z,k)(i,j)}$, $z{\neq}k, z\in(1,...,I)$ in the other groups.
\item
Now, inside each group $(x,y)$ processor $P_{(x,y)(i,j)}$ has the required part of the pivot row of the matrix $B$ and it further vertically broadcast it to the processors $P_{(x,y)(r,j)}$,
$r{\neq}j, r\in(1,...,\frac{t}{J})$ inside the group.
\end{itemize}
\item
Each processor inside a group updates each block in its $C$ rectangle with one block from the pivot column and one block from the pivot row,
so that each block $c_{ij}, (i, j)\in(1,...,\frac{n}{b})$ of matrix $C$ will be updated as $c_{ij}=c_{ij}+a_{ik}{\times}{b_{kj}}$.
\item
After $\frac{n}{b}$ steps of the algorithm, each block $c_{ij}$ of matrix $C$ will be equal to $\sum_{k=1}^{\frac{n}{b}}a_{ik}\times{b_{kj}}$
\end{itemize}

\begin{figure}[!t]
\begin{center}
\begin{tikzpicture}[transform shape, scale=0.7]
  \draw (0, 0) grid (6, 6);
  \draw[very thick] (0, 0) grid[step=2] (6, 6);
  \draw[pattern=north west lines] (0.5, 0) rectangle (1, 6);
  \draw[pattern=north east lines] (0,5.5) rectangle (6, 5);
  \draw[->,double distance=1pt] (3.5,5.25) .. controls ++(0.25,-0.5) and
  ++(0.25,0.5).. (3.5, 3.25);
  \draw[->,double distance=1pt] (3.5,5.25) .. controls ++(0.5,-0.5) and
  ++(0.5,0.5).. (3.5, 1.25);
  \draw[->,double distance=1pt] (0.75,3.5) .. controls ++(0.5,0.25) and
  ++(-0.5,0.25).. (2.75, 3.5);
  \draw[->,double distance=1pt] (0.75,3.5) .. controls ++(0.5,0.5) and
  ++(-0.5,0.5).. (4.75, 3.5);
\end{tikzpicture}
\end{center}
\caption{HSUMMA between groups}
\label{hsumma:step1}
\end{figure}
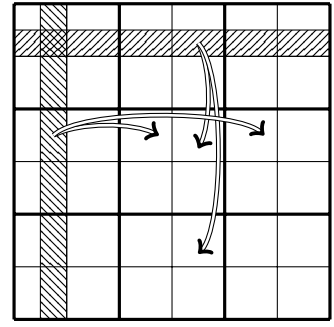

\begin{figure}[!t]
\begin{center}
\begin{tikzpicture}[transform shape, scale=0.7]
  \draw[very thick] (0, 0) grid[step=3] (6, 6);
  \draw[pattern=north west lines] (1.5, 0) rectangle (3, 6);
  \draw[pattern=north east lines] (0,4.5) rectangle (6, 3);
  \foreach \n in {1.5,1.875,...,2.6725}{
    \fill ($ (\n,6-\n)$) rectangle ++(0.375,-0.375);
  }
  \node[text=white, font=\large] at (1.7, 4.3) {\textbf{b}};
  \node[text=black, font=\large] at (5, 3.8) {\textbf{B}};
  \node[text=black, font=\large] at (2.2, 1.5) {\textbf{B}};
\end{tikzpicture}
\end{center}
\caption{HSUMMA inside group}
\label{hsumma:step2}
\end{figure}
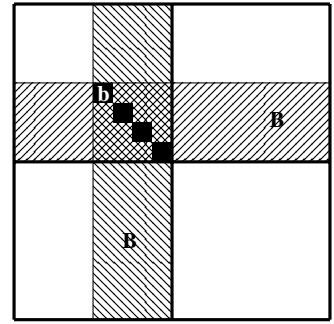

The communication phases described above are illustrated by Figure~\ref{hsumma:step1}.
In general it is possible to use one block size between groups and another block size inside a group.
In this case the size of sent data between the groups is at least the same as the size of data sent inside a group.
Therefore, the block size inside a group should be less than or equal to the block size between groups.
Let us assume the block size between groups is $B$ and inside a group is $b$. Then, the number of steps in the higher level
will be equal to the number of blocks between groups: $\frac{n}{B}$. In each iteration between the groups,
the number of steps inside a group is $\frac{B}{b}$, so the total number of steps of HSUMMA, $\frac{n}{B}{\times}\frac{B}{b}$, will be the same as the number of steps of SUMMA.
The amount of data sent is the same as in SUMMA. The steps inside a group are shown in Figure~\ref{hsumma:step2}.
It is clear that SUMMA is a special case of HSUMMA when the number of groups equals to one or to the total number of processors.

One may ask why to introduce a new hierarchical algorithm if MPI implementations already provide us
with high-performance broadcast algorithms. As we have already mentioned before, MPI optimizations of broadcast are
platform specific and do not depend on the application. On the other hand, HSUMMA is an application specific hierarchical algorithm 
which optimizes communication cost at the application level and is platform independent. A general purpose broadcast algorithm can not replace
the communication pattern of HSUMMA as in each level it calculates pivot row and pivot column before broadcasting and it is very application specific.

The pseudocode for HSUMMA is Algorithm~\ref{Alg:Hierarchical}.
\begin{algorithm}
\DontPrintSemicolon
\CommentSty{/*The A,B,C matrices are distributed on a virtual 2-D grid of
  $p=s{\times}t$ processors.}\;
\CommentSty{Here are the instructions executed by the processor
  $\text{P}_{\text{(x,y)}\text{(i,j)}}$ (this is the processor (i,j)
  inside the group (x,y)).*/}\;
 \KwData{$\text{NB}_{\text{Block}\_\text{Group}}$: Number of steps in the higher level}
 \KwData{$\text{NB}_{\text{Block}\_\text{Inside}}$: Number of steps in the lower level}
 \KwData{$(M,L,N)$: Matrix dimensions}
 \KwData{$A, B$: two input sub-matrices
 of size $(\frac{M}{s} \times \frac{L}{t},
           \frac{L}{s} \times \frac{N}{t})$}
 \KwResult{$C$: result sub-matrix of size $\frac{M}{s}
   \times \frac{N}{t}$}
\Begin{
\tcc*[l]{Broadcast A and B}
\MPIC group\_col\_comm \tcc*[r]{communicator between $\text{P}_{(*,y)(i,j)}$ processors}
\MPIC group\_row\_comm \tcc*[r]{communicator between $\text{P}_{(x,*)(i,j)}$ processors}
\MPIC col\_comm \tcc*[r]{communicator between $\text{P}_{(x,y)(*,j)}$ processors}
\MPIC row\_comm \tcc*[r]{communicator between $\text{P}_{(x,y)(i,*)}$
  processors}\;

\For {$\text{iter}_\text{group} = 0$; $\text{iter}_\text{group} <
  \text{NB}_{\text{Block}\_{\text{Group}}}$;
  $\text{iter}_\text{group}++$}{\;
\If{i == Pivot\_inside\_group\_col(iter$_\text{group}$)}{
\If{x == Pivot\_group\_col(iter$_\text{group}$)}{
  \;\tcc*[l]{Get direct access to the iter$_\text{group}^\text{th}$
    group block of A}
  Copy\_Block\_group( Block$_{\text{group}\_{\text{A}}}$, A,
                      iter$_\text{group}$ )\;
}
  \MPIB{Block$_{\text{group}\_{\text{A}}}$,
             Type$_{\text{Block}\_{\text{group\_A}}}$,
             Pivot\_group\_col(iter$_\text{group}$), group\_row\_comm }\;
}\;
\If{j == Pivot\_inside\_group\_row(iter$_\text{group}$)}{
\If{y == Pivot\_group\_row(iter$_\text{group}$)}{
  \;\tcc*[l]{Get direct access to the iter$_\text{group}^\text{th}$
    group block of B}
  Copy\_Block\_group( Block$_{\text{group}\_\text{B}}$, B, iter$_\text{group}$ )\;
}
  \MPIB{Block$_{\text{group}\_\text{B}}$,
             Type$_{\text{Block}\_{\text{group\_B}}}$,
             Pivot\_group\_row(iter$_\text{group}$),
             group\_col\_comm }\;
}\;
\For {$\text{iter} = 0$; $\text{iter} <
  \text{NB}_{\text{Block}\_\text{Inside}}$; $\text{iter}++$}{
\If{i == Pivot\_inside\_group\_col(iter)}{
  \;\tcc*[l]{Get access to the iter$^\text{th}$ block of
    Block$_{\text{group}\_{\text{A}}}$ on this processor}
  \CPBA{ Block$_\text{A}$, Block$_{\text{group}\_{\text{A}}}$, iter }\;
}
  \MPIB{Block$_\text{A}$, Type$_\text{Block\_A}$, \Pcol{iter}, row\_comm }\;\;
\If{j == Pivot\_inside\_group\_row(iter)}{
  \;\tcc*[l]{Get access to the iter$^\text{th}$ block of
    Block$_{\text{group}\_{\text{B}}}$ on this processor}
  \CPBB{ Block$_\text{B}$, Block$_{\text{group}\_{\text{B}}}$, iter }\;
}
  \MPIB{Block$_\text{B}$, Type$_\text{Block\_B}$, \Prow{iter},
             col\_comm }\;\;
  \DGemm{Block$_\text{A}$, Block$_\text{B}$, C}\;
}
}
}

\caption{Hierarchical SUMMA algorithm}
\label{Alg:Hierarchical}
\end{algorithm}
\section{Theoretical Analysis}
\label{Sec:theory}
In this section SUMMA and HSUMMA are theoretically analysed and compared.
First of all, for simplicity we assume that the matrices are $n\times{n}$ square matrices. Let $b$ be block size inside one group 
and $B$ be block size between groups.
The execution time depends on the communication time (i.e. the broadcast algorithm and the communication model).
As a communication model we use Hockney's model~\cite{Hockney:1994:CCM:179977.179986} which represents the time of sending of
a message of size $m$ between two processors as $\alpha+m\beta$. Here, $\alpha$ is the latency, $\beta$ is the reciprocal
of network bandwidth. In addition, let us assume that a combined floating point computation(for one addition and multiplication) time is $\gamma$.
Binomial tree and Van de Geijn broadcast algorithms \cite{Barnett94interprocessorcollective}\cite{Shroff00collmark:mpi} are used to analyse both our algorithm and SUMMA. 
It is known that the costs of these broadcast algorithms are as follows:
\begin{itemize}
\item
Binomial tree algorithm: $\log_2(p)\times{(\alpha+m\times{\beta}})$
\item
Van de Geijn algorithm: $(\log_2(p)+p-1)\alpha+2\frac{p-1}{p}m\beta$
\end{itemize}
\subsection{Analysis of SUMMA}
For simplicity let us assume the $n\times{n}$ matrices are distributed over a two-dimensional $\sqrt{p}{\times}\sqrt{p}$ grid of
processors and the block size is $b$. This algorithm has $\frac{n}{b}$ steps. In each step, processors broadcast pivot row and pivot column.
So the computation cost in each step is
$O(2\times{\frac{n^2}{p}{\times}{b}})$. 
Hence, the overall computation cost will be $O(\frac{2n^3}{p})$.

The broadcasts of each row and column are independent at each step and they can be done in parallel.
For this analysis the network congestion is neglected. The amount of data transferred by each broadcast is $\frac{n}{\sqrt{p}}\times{b}$.
The total communication cost of SUMMA can be computed by multiplying the
communication cost of each step by the number of steps depending on the broadcast algorithm.
The results are:
\begin{itemize}
\item
Binomial Tree: \newline $\log_2{\left(p\right)}{\times} \left(\alpha\frac{n}{b} + \beta{\times} \frac{ n^2}{\sqrt{p}}\right)$
\item
Van de Geijn broadcast: \newline $(\log_2{\left(p\right)} + 2(\sqrt{p} - 1))\alpha\frac{n}{b} + 4(1-\frac{1}{\sqrt{p}})\frac{ n^2}{\sqrt{p}}\beta$ 
\end{itemize}

\subsection{Analysis of HSUMMA}
To simplify the analysis, let us assume there are $G$ groups arranged as $\sqrt{G}\times{\sqrt{G}}$ grid of processors groups.
Let $B$ denote the block size between groups(we also call such a block an outer block), $b$ be block size inside a group, and $n{\times}n$ be the size of the matrices.

There is one step per outer block, thus there will be $\frac{n}{B}$ steps in the highest level called outer steps.
Each outer block belongs to $\sqrt{p}$ processors. These processors broadcast the part of the outer block along the row or column of
processor in parallel. The size of sent data is $2\frac{n\times{B}}{\sqrt{p}}$ per processor which has a part of the outer block.

Inside one group, processors are arranged in a grid of size:
$\frac{\sqrt{p}}{\sqrt{G}}\times{\frac{\sqrt{p}}{\sqrt{G}}}$.
After the reception of the outer block each group multiplies the outer block using the SUMMA algorithm inside the group.
Thus there are $\frac{B}{b}$ inner steps to execute. The inner block belongs to $\frac{\sqrt{p}}{\sqrt{G}}$ processors.
These processors send $2\frac{n\times{b}}{\sqrt{p}}$ amount of data per inner step.

The overall communication time is equal to the sum of the communication times between the groups and inside the groups.
\begin{itemize}
\item
Inner Communication cost (inside each group):
\begin{itemize}
\item
Binomial Tree: \newline
$\log_2{\left(\frac{p}{G}\right)}{\times}\left(\alpha{\times}\frac{n}{b}+\beta {\times}\frac{n^2}{\sqrt{p}}\right)$
\item
Van de Geijn broadcast: \newline
$\left(\log_2{\left(\frac{p}{G}\right)}+2\left(\frac{\sqrt{p}}{\sqrt{G}}-1\right)\right){\times}\alpha{\times} \frac{n}{ b}+4(1-\frac{\sqrt{G}}{\sqrt{p}}){\times}\frac{n^2}{\sqrt{p}}\beta $
\end{itemize}
\item
Outer Communication cost (between groups):
\begin{itemize}
\item Binomial Tree:
$\log_2{\left(G\right)}{\times}\left(\alpha{\times}\frac{n}{B}+\beta{\times}\frac{n^2}{\sqrt{p}}\right) $
\item
Van de Geijn broadcast: \newline
$\left(\log_2{\left(G\right)}+2\left(\sqrt{G} -1\right)\right){\times}\alpha{\times}\frac{n}{B}+4(1-\frac{1}{\sqrt{G}}){\times}\frac{n^2}{\sqrt{p}}\beta $
\end{itemize}
\end{itemize}
We can summarize the cost of HSUMMA and SUMMA as in Table~\ref{comp:binomial} and Table~\ref{comp:geijn}.
\begin{table}[H]
\centering
\caption{Comparison with binomial tree broadcast}
\scalebox{0.6}{
\begin{tabular}{|c|c|c|c|c|c|}
  \hline
&&\multicolumn{2}{c|}{}&\multicolumn{2}{c|}{}\\
Algorithm & Comp. Cost&\multicolumn{2}{c|}{Latency Factor}&\multicolumn{2}{c|}{Bandwidth Factor}\\
\cline{3-6}
&&inside groups&between groups&inside groups&between groups\\
  \hline
&&\multicolumn{2}{c|}{}&\multicolumn{2}{c|}{}\\
SUMMA&$\frac{2n^3}{p}$&\multicolumn{2}{c|}{
$\log_2{\left(p\right)}{\times}\frac{n}{b}$}&\multicolumn{2}{c|}{
$n^2{\times}\frac{\log_2{\left(p\right)}}{\sqrt{p}}$}\\
&&\multicolumn{2}{c|}{}&\multicolumn{2}{c|}{}\\
  \hline
&&&&&\\
HSUMMA&$\frac{2n^3}{p}$&
$\log_2{\left(\frac{p}{G}\right)}{\times}\frac{n}{b}$&
$\log_2{\left(G\right)}{\times} \frac{n}{B}$&
$\log_2{\left(\frac{p}{G}\right)}{\times}\frac{n^2}{\sqrt{p}}$&
$\log_2{\left(G\right)}{\times}\frac{n^2}{\sqrt{p}}$\\
&&&&&\\
  \hline
\end{tabular}}
\label{comp:binomial}
\end{table}

\begin{table*}[!t]
\centering
\caption{Comparison with Van De Geijn broadcast}
\scalebox{0.8}{
\begin{tabular}{|c|c|c|c|c|c|}
  \hline
&&\multicolumn{2}{c|}{}&\multicolumn{2}{c|}{}\\
Algorithm & Comp. Cost&\multicolumn{2}{c|}{Latency Factor}&\multicolumn{2}{c|}{Bandwidth Factor}\\
\cline{3-6}
&&inside groups&between groups&inside groups&between groups\\
  \hline
&&\multicolumn{2}{c|}{}&\multicolumn{2}{c|}{}\\
SUMMA&$\frac{2n^3}{p}$&\multicolumn{2}{c|}{
$\left(\log_2{\left(p\right)} + 2\left(\sqrt{p} - 1\right)\right){\times}\frac{n}{b}$}&\multicolumn{2}{c|}{
$4\left(1-\frac{1}{\sqrt{p}}\right) {\times} \frac{n^2}{\sqrt{p}}$}\\
&&\multicolumn{2}{c|}{}&\multicolumn{2}{c|}{}\\
  \hline
&&&&&\\
HSUMMA&$\frac{2n^3}{p}$&
$\left(\log_2{\left(\frac{p}{G}\right)} + 2\left(\frac{\sqrt{p}}{\sqrt{G}} - 1\right)\right){\times}\frac{n}{b}$&
$\left(\log_2{\left(G\right)} + 2\left(\sqrt{G} -1\right)\right){\times} \frac{n}{B}$&
$4\left(1-\frac{\sqrt{G}}{\sqrt{p}}\right){\times}\frac{n^2}{\sqrt{p}}$&
$4\left(1-\frac{1}{\sqrt{G}}\right){\times}\frac{n^2}{\sqrt{p}}$\\
&&&&&\\
  \hline
&&\multicolumn{2}{c|}{}&\multicolumn{2}{c|}{}\\
HSUMMA($G=\sqrt{p}$, $b=B$) &$\frac{2n^3}{p}$&\multicolumn{2}{c|}{
$\left(\log_2{\left(p\right)} + 4\left(\sqrt[4]{p} - 1\right)\right){\times}\frac{n}{b}$}&\multicolumn{2}{c|}{
$8\left(1-\frac{1}{\sqrt[4]{p}}\right) {\times} \frac{n^2}{\sqrt{p}}$}\\
&&\multicolumn{2}{c|}{}&\multicolumn{2}{c|}{}\\
  \hline
\end{tabular}}
\label{comp:geijn}
\end{table*}

\subsection{Theoretical Prediction}
\label{Sec:general_model}
One of the goals of this section is to demonstrate that independent on the broadcast algorithm employed by SUMMA, 
HSUMMA will either outperform SUMMA or be at least equally fast.
In this section we introduce a general model for broadcast algorithms and theoretically predict SUMMA and HSUMMA.
In the model we assume no contention and assume all the links are homogeneous. We prove that even with this simple model 
the extremums of the communication cost function can be predicted.

Again we assume that the time taken to send a message of size $m$ between any two processors is modeled as $T(m)=\alpha+m{\times}\beta$, where
$\alpha$ is the latency and $\beta$ is the reciprocal bandwidth.

We model a broadcast time for a message of size $m$ among $p$ processors by formula (\ref{eq:bcast}). This model 
generalizes all homogeneous broadcast algorithms such as pipelined/non-pipelined flat, binary, binomial, linear, scatter/gather broadcast 
algorithms\cite{Grebovic:2007}\cite{Patarasuk:2008:TPB:1370314.1370586} which are used inside state of the art broadcast implementations like MPICH and Open MPI.
\begin{equation}
T_{bcast}(m,p)=L(p){\times}\alpha+m{\times}W(p){\times}\beta
\label{eq:bcast}
\end{equation}
In (\ref{eq:bcast}) we assume that $L(1)=0$ and $W(1)=0$. It is also assumed that $L(p)$ and $W(p)$ are 
monotonic and differentiable functions in the interval $(1,p)$ and their first derivatives are constants or monotonic in the interval $(1,p)$.

With this general model the theoretical communication cost of SUMMA will be as follows:
\begin{equation}
T_{S}(n,p)=2\left(\frac{n}{b}{\times}L(\sqrt{p})\alpha+\frac{n^2}{\sqrt{p}}{\times}W(\sqrt{p})\beta \right)
\label{eq:summa}
\end{equation}
In the same way we can express the communication cost of HSUMMA as the sum of the latency cost and the bandwidth cost:
\begin{equation}
T_{HS}(n,p,G)=T_{HS_{l}}(n,p,G) + T_{HS_{b}}(n,p,G)
\label{eq:hsumma}
\end{equation}
Here $G\in[1,p]$ and we take $b=B$ for simplicity. The latency cost $T_{HS_{l}}(n,p,G)$ and the bandwidth cost $T_{HS_{b}}(n,p,G)$
will be given by the following formulas:
\begin{equation}
T_{HS_{l}}(n,p,G)=2\frac{n}{b}{\times}\left (L(\sqrt{G})+L(\frac{\sqrt{p}}{\sqrt{G}})\right)\alpha
\label{eq:hsumma_lat}
\end{equation}
\begin{equation}
T_{HS_{b}}(n,p,G)=2\frac{n^2}{\sqrt{p}} {\times} \left( W(\sqrt{G})+W(\frac{\sqrt{p}}{\sqrt{G}})  \right)\beta
\label{eq:hsumma_band}
\end{equation}
It is clear that $T_{S}(n,p)$ is a speacial case of $T_{HS}(n,p,G)$ when $G=1$ or $G=p$.

Let us investigate extremums of $T_{HS}$ as a function of $G$ for a fixed $p$ and $n$. We have $b=B$. 
\begin{equation}
\frac{{\partial}T_{HS}}{{\partial}G}=\frac{n}{b}{\times}L_1(p,G)\alpha+\frac{n^2}{\sqrt{p}}{\times}W_1(p,G)\beta
\end{equation}
Here, $L_1(p,G)$ and $W_1(p,G)$ are defined as follows:
\begin{equation}
L_1(p,G)=\left(\frac{{\partial}L(\sqrt{G})}{{\partial}\sqrt{G}}{\times}\frac{1}{\sqrt{G}} 
- \frac{{\partial}L(\frac{\sqrt{p}}{\sqrt{G}})}{{\partial}\frac{\sqrt{p}}{\sqrt{G}}}{\times}\frac{\sqrt{p}}{G\sqrt{G}}\right)
\label{eq:hsumma_lat_der}
\end{equation}
\begin{equation}
W_1(p,G)=\left(\frac{{\partial}W(\sqrt{G})}{{\partial}\sqrt{G}}{\times}\frac{1}{\sqrt{G}} 
- \frac{{\partial}W(\frac{\sqrt{p}}{\sqrt{G}})}{{\partial}\frac{\sqrt{p}}{\sqrt{G}}}{\times}\frac{\sqrt{p}}{G\sqrt{G}}\right)
\label{eq:hsumma_band_der}
\end{equation}

It can be easily shown that, if $G=\sqrt{p}$ then $L_1(p,G)=0$ and $W_1(p,G)=0$, thus, $\frac{{\partial}T_{HS}}{{\partial}G}=0$. In addition, 
$\frac{{\partial}T_{HS}}{{\partial}G}$ changes the sign in the interval $(1,p)$ depending on the value of $G$. 
That means $T_{HS}(n,p,G)$ has extremum at $G=\sqrt{p}$ for fixed $n$ and $p$. 
The expression of $\frac{{\partial}T_{HS}}{{\partial}G}$ shows that, depending on the ratio of $\alpha$ and $\beta$ the extremum can be either minimum or 
maximum in the interval $(1,p)$. If $G=\sqrt{p}$ is the minimum point it means that with $G=\sqrt{p}$ HSUMMA will outperform SUMMA, 
otherwise HSUMMA with $G=1$ or $G=p$ will have the same performance as SUMMA. 

Now lets apply this analysis to the HSUMMA communication cost function obtained for Van de Geijn broadcast algorithm (see Table~\ref{comp:geijn})
and again assuming $b=B$ for simplicity. We will have:
\begin{equation}
\frac{{\partial}T_{HS_V}}{{\partial}G} = \frac{G-\sqrt{p}}{G\sqrt{G}}{\times}\left(\frac{n\alpha}{b} - 2\frac{n^2}{p}{\times}\beta\right)
\label{eq:hsumma_vg}
\end{equation}

It is clear that if $G=\sqrt{p}$ then $\frac{\partial{T_{HS_V}}}{\partial{G}}=0$. 
Depending on the ratio of $\alpha$ and $\beta$, the communication cost as a function of $G$ has either minimum or maximum in the interval $(1,p)$.
\begin{itemize}
\item
If 
\begin{equation}
\frac{\alpha}{\beta} > 2\frac{nb}{p}
\label{eq:ratio}
\end{equation}
then $\frac{\partial{T_{HS_V}}}{\partial{G}} < 0$ in the interval $(1,\sqrt{p})$ and $\frac{\partial{T_{HS_V}}}{\partial{G}} > 0$ in $(\sqrt{p},p)$.
Thus $T_{HS}$ has the minimum in the interval $(1,p)$ and the minimum point is $G=\sqrt{p}$. 
\item
If 
\begin{equation}
\frac{\alpha}{\beta} < 2\frac{nb}{p}
\end{equation}
then $T_{HS}$ has the maximum in the interval $(1,p)$ and the maximum point is $G=\sqrt{p}$. The function gets its minimum at
either $G=1$ or $G=p$. 
\end{itemize}

If we take $G=\sqrt{p}$ in the HSUMMA communication cost function (see Table~\ref{comp:geijn}) and assume the above conditions the optimal communication 
cost function will be as follows:
\begin{equation}
\left(\log_2{\left(p\right)} + 4\left(\sqrt[4]{p} - 1\right)\right){\times}\frac{n}{b}{\times}\alpha + 8\left(1-\frac{1}{\sqrt[4]{p}}\right) {\times} \frac{n^2}{\sqrt{p}}{\times}\beta
\label{eq:hsumma_optimal}
\end{equation}
Thus, we have proved that depending on the ratio of $\alpha$ and $\beta$ HSUMMA will either reduce the communication cost of SUMMA
or in the worst case have the same performance as SUMMA.

We will use this model to predict the performance of HSUMMA on Grid5000, BlueGene/P and future exascale platforms.
\section{Experiments}
\label{Sec:experiments}
Our experiments were carried out on a cluster of Grid5000 and a BlueGene/P (BG/P) platform which are fairly representative and span a good spectrum of
loosely and tightly coupled platforms. The details of the platforms are given in the appropriate sections. 
The times in our experimental results are the mean times of $30$ experiments. 
\subsection{Experiments on Grid5000}
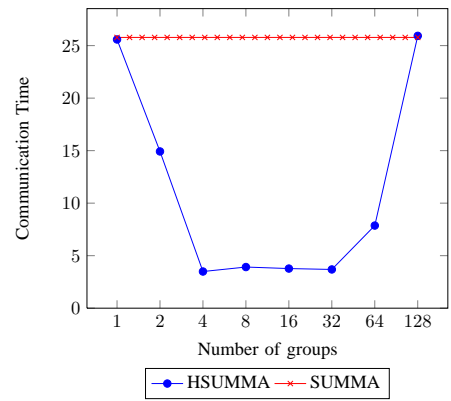
\begin{figure}[H]
\centering
\begin{tikzpicture}[scale=0.7]
\begin{semilogxaxis}[
        xlabel=Number of groups,
        ylabel=Communication Time,
        log basis x=2,
        legend style={at={(0.5,-0.2)},anchor=north,legend columns=2},
        ymin = 0,
        xticklabel=\pgfmathparse{2^\tick}\pgfmathprintnumber{\pgfmathresult}
        ]
\addplot[color=blue, mark=*] table[x="NB_groups",y="overall_comm"]
{data/g5000_hsumma_p128_N8192_b64.dat};
\addplot[red, domain=1:128, mark=x] {25.78};
\legend{HSUMMA, SUMMA}
\end{semilogxaxis}
\end{tikzpicture}
\caption{HSUMMA on Grid5000. Communication time in seconds. $b=B=64$,$n=8192$ and $p=128$} 
\label{fig:g5000_p128_N8192_hsumma_b64_B64_groups_comm}
\end{figure}

\begin{figure}[H]
\centering
\begin{tikzpicture}[scale=0.7]
\begin{semilogxaxis}[
        xlabel=Number of groups,
        ylabel=Communication Time,
        log basis x=2, 
	legend style={at={(0.5,-0.2)},anchor=north,legend columns=2},
        ymin = 0,
        xticklabel=\pgfmathparse{2^\tick}\pgfmathprintnumber{\pgfmathresult}
        ]
\addplot[color=blue, mark=*] table[x="NB_groups",y="overall_comm"]
{data/g5000_hsumma_p128_N8192_b512.dat};
\addplot[red, domain=1:128, mark=x] {4.53};
\legend{HSUMMA, SUMMA}
\end{semilogxaxis}
\end{tikzpicture}
\caption{HSUMMA on Grid5000. Communication time in seconds. $b=B=512$, $n=8192$ and $p=128$}
\label{fig:g5000_p128_N8192_hsumma_b512_B512_groups_comm}
\end{figure}
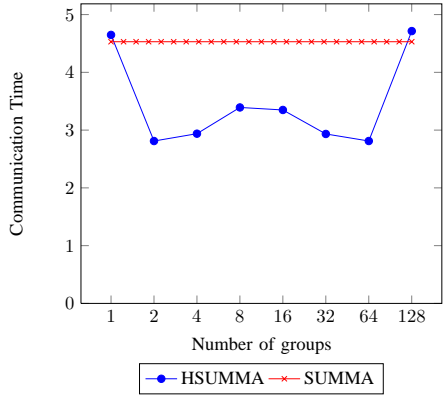

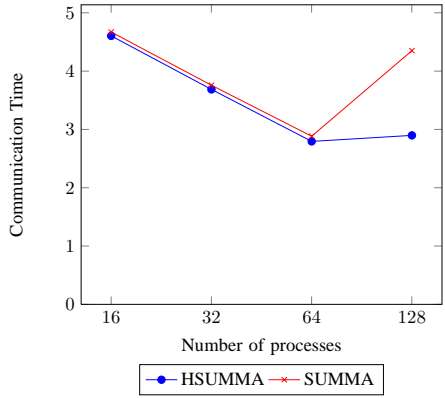
\begin{figure}[H]
\centering
\begin{tikzpicture}[scale=0.7]
\begin{semilogxaxis}[
        xlabel=Number of processes,
        ylabel=Communication Time,
        log basis x=2,
	legend style={at={(0.5,-0.2)},anchor=north,legend columns=2},
        ymin = 0,
        ytick={0,1,2,3,4,5,6,7,8},
        xticklabel=\pgfmathparse{2^\tick}\pgfmathprintnumber{\pgfmathresult}
        ]
\addplot[color=blue, mark=*] table[x="NB_procs",y="overall_comm"]
{data/g5000_hsumma_N8192_b512_procs.dat};
\addplot[color=red, mark=x] table[x="NB_proc",y="comm_mean"]
{data/g5000_csumma_N8192_b512_procs.dat};
\legend{HSUMMA, SUMMA}
\end{semilogxaxis}
\end{tikzpicture}
\caption{HSUMMA and SUMMA on Grid5000. Communication time in seconds. $b=B=512$ and $n=8192$}
\label{fig:g5000_N8192_hsumma_b512_B512_procs_comm}
\end{figure}

Some of our experiments were carried out on the Graphene cluster of Nancy site of Grid5000 platform.
We have used Intel MKL BLAS for sequential operations, MPICH-2 for MPI implementation and our implementations of the matrix multiplication algorithms.
In addition to MPICH we also did some experiments with Open MPI on Grid5000 and got similar results. Thus in this paper we just present the experiments with
MPICH implementation of MPI. The size of matrices in our experiments on Grid5000 is $8192{\times}8192$. 
Figure~\ref{fig:g5000_p128_N8192_hsumma_b64_B64_groups_comm} compares SUMMA and HSUMMA with block size $64$. It is clear that
smaller block sizes lead to a larger number of steps and this in turn will affect the latency cost. It can be seen that in this case HSUMMA outperforms SUMMA
with huge difference. 
Figure~\ref{fig:g5000_p128_N8192_hsumma_b512_B512_groups_comm} represents the same comparision but with block size $512$. This block size
is the maximum possible one with this configuration. In this case the improvement is $1.6$ times as the minimum communication time of HSUMMA
and SUMMA are $2.81$ and $4.53$ seconds respectively. In addition, theoretically HSUMMA should has the same performance as SUMMA when $G=1$ or $G=p$
and the figures verifies that in practice. That means HSUMMA can never be worse than than SUMMA. In the worst case it will have the same performance
as SUMMA.
Figure~\ref{fig:g5000_N8192_hsumma_b512_B512_procs_comm} shows experimental results from scalability point of view.
Here we use the largest possible block size for both algorithms. If we used block size $64$ for scalability plot we would see the 
significant difference between HSUMMA and SUMMA. However, even with this configuration which is optimal for SUMMA it can be seen that on small platforms both SUMMA and 
HSUMMA have the same performance, however, the trend shows that on larger platforms HSUMMA will outperform SUMMA and therefore is more scalable.
The experiments show that with any number of groups HSUMMA outperforms SUMMA on Grid5000. 

\subsubsection{Validation of the Anlytical Model on Grid5000}
We take the following approximately real parameters for Graphene cluster of Grid5000:
\begin{itemize}
\item
Latency: 1E-4
\item
Reciprocal bandwidth: 1E-9
\item
p: 8192
\item
n: 8192
\item
b: 64
\end{itemize}
The algorithmic parameters are the same as in our experiments on Grid5000.
It is clear that $\frac{\alpha}{\beta} > 2*\frac{8192*64}{128}=8192$ and therefore according to our theoretical analysis HSUMMA has minimum in the interval $(1,p)$.
We do not have experimental minimum exactly at $G=\sqrt{p}$ as predicted by our theoretical results. However, this does not downgrade the importance of our analytical 
model because the main goal of our analytical analysis is to predict if HSUMMA will be more efficient than SUMMA on the target platform or not.
If this is the case, the optimal number of groups can be easily found experimentally by using only few iterations of HSUMMA with different values of $G$ 
and thus can be incorporated into the algorithm.

\subsection{Experiments on BlueGene/P}
Some of our experiments were carried out on Shaheen BlueGene/P at Supercomputing Laboratory at King Abdullah University
of Science\&Technology (KAUST) in Thuwal, Saudi Arabia. Shaheen is a 16-rack BlueGene/P. Each node is equipped with four 32-bit, 850 Mhz PowerPC 450 cores and 4GB DDR memory. 
VN (Virtual Node) mode with torus connection was used for the experiments on the BG/P. The Blue Gene/P architecture provides a three-dimensional point-to-point Blue Gene/P torus network 
which interconnects all compute nodes and global networks for collective and interrupt operations. Use of this network is integrated into the BlueGene/P MPI implementation.

All the sequential computations in our experiments were performed by using DGEMM routine from IBM ESSL library. 
The size of the matrices for all our experiments on the BG/P is \textbf{$65536{\times}65536$}.
We use our implementation of SUMMA for comparison with HSUMMA as the performance of ScaLAPACK implementation lingers behind our implementation.

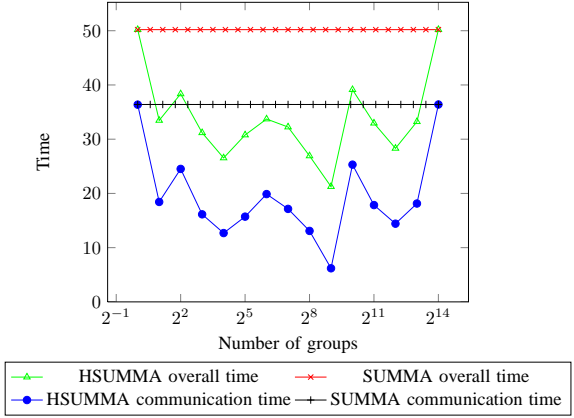
\begin{figure}[H]
\centering
\begin{tikzpicture}[scale=0.7]
\begin{semilogxaxis}[
        xlabel=Number of groups,
        ylabel=Time,
        log basis x=2,
        legend style={at={(0.5,-0.2)},anchor=north,legend columns=2},
        ymin = 0
        ]
\addplot[color=green, mark=triangle] table[x="NB_groups",y="time_mean"]
{data/bg_p16384_nc.dat};
\addplot[red, domain=1:16384, mark=x] {50.2};
\addplot[color=blue, mark=*] table[x="NB_groups",y="overall_comm"]
{data/bg_p16384_nc.dat};
\addplot[black, domain=1:16384, mark=+] {36.4};
\legend{HSUMMA overall time, SUMMA overall time, HSUMMA communication time, SUMMA communication time}
\end{semilogxaxis}
\end{tikzpicture}
\caption{SUMMA and HSUMMA on 16384 cores on BG/P. Execution and communication time. $b=B=256$ and $n=65536$}
\label{fig:bg_p16384_N32768_hsumma_b256_groups_comm_test}
\end{figure}

The benefit of HSUMMA comes from the optimal number of groups. Therefore, it is interesting to see how different numbers of groups affect 
the communication cost of HSUMMA on a large platform. Figure~\ref{fig:bg_p16384_N32768_hsumma_b256_groups_comm_test} shows HSUMMA on $16384$ cores.
In order to have a fair comparison we use the same block size inside a group and between the groups.
The figure shows that the execution time of SUMMA is $50.2$ seconds and the communication time is $36.46$ seconds. 
On the other hand, the minimum execution time of HSUMMA is $21.26$ and the minimum communication time is $6.19$ when $G=512$.
Thus the execution time of HSUMMA is $2.36$ times and the communication time is $5.89$ times less than that of SUMMA on $16384$
cores. On the other hand, HSUMMA achieves $2.08$ times less communication time and $1.2$ times less overall execution time than SUMMA on $2048$ cores.
We also did experiments on BlueGene/P cores smaller than $2048$ and the results showed that on smaller numbers of cores the performance of HSUMMA and
SUMMA was almost the same. 

The "zigzags" on the figure can be explained by the fact that mapping communication layouts to network hardware on BlueGene/P impacts 
the communication performance and it was observed by P. Balaji et al.\cite{Balaji:mpibg} as well. When we group processors we do not take into account the platform parameters. 
However, according to our preliminary observations these "zigzags" can be eliminated by taking platform parameters into account while grouping. In addition, the effects of
square versus non-square meshes also a reason for that. 

\begin{figure}[H]
\centering
\begin{tikzpicture}[scale=0.7]
\begin{semilogxaxis}[
        xlabel=Number of procs,
        ylabel=Communication Time,
        log basis x=2,
        legend style={at={(0.5,-0.2)},anchor=north,legend columns=2},
        ymin = 0,
        ]
\addplot[color=blue, mark=*] table[x="NB_procs",y="overall_comm"]
{data/bg_n65536_hsumma_vn_procs.dat};
\addplot[color=red, mark=square] table[x="NB_procs",y="overall_comm"]
{data/bg_n65536_summa_vn_procs.dat};
\legend{HSUMMA communication time, SUMMA communication time}
\end{semilogxaxis}
\end{tikzpicture}
\caption{HSUMMA and SUMMA on BlueGene/P with VN mode. Communication time in seconds. $b=B=256$ and $n=65536$}
\label{fig:bg_hsumma_summa_procs_comm_vn}
\end{figure}
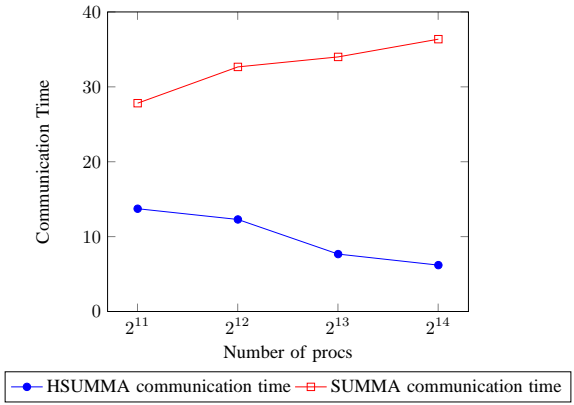

Figure~\ref{fig:bg_hsumma_summa_procs_comm_vn} represents scalability analysis of SUMMA and HSUMMA from communication 
point of view. 
It can be seen that HSUMMA is more scalable than SUMMA and this pattern suggests that the communication performance of HSUMMA 
gets much better than that of SUMMA while the number of cores increases.

\subsubsection{Validation of the Anlytical Model on BlueGene/P}
The parameters of the BlueGene/P are as follows: 
\begin{itemize}
\item
Latency: 3E-6
\item
Bandwidth: 1E-9
\item
p: 16384
\item
n: 65536
\item
b: 256
\end{itemize}
Here again we use the same values of the algorithmic parameters as in our experiments.
By using these values it can be shown that $\frac{\alpha}{\beta}>2\frac{nb}{p}$ which proves the communication function of HSUMMA has the minimum in the interval $(1,p)$.
For some ratios of $n$ and $p$ the above condition may not hold. However, in this case the cost of matrix multiplication will be dominated by
computation cost and even in this case HSUMMA can be used just by using one or $p$ group.
\subsection{Prediction on Exascale}
We use the following parameters to predict performance of HSUMMA on exascale platforms. These platform parameters are obtained from a recent report on
exascale architecture roadmap\cite{exascale:report}.
\begin{itemize}
\item
Total flop rate ($\gamma$): $1E18$ flops
\item
Latency: $500$ ns
\item
Bandwidth: $100$ GB/s
\item
Problem size: $n=2^{22}$
\item
Number of processors: $p=2^{20}$
\item
Block size: $b=256$ 
\end{itemize}

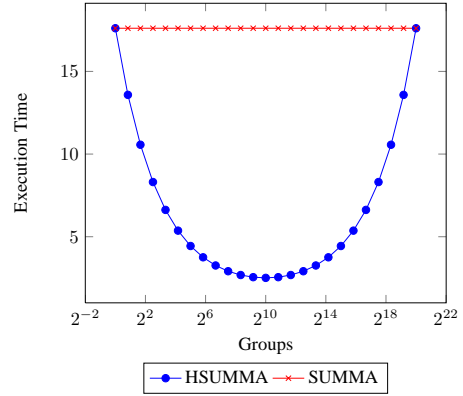
\begin{figure}[H]
\centering
\newcommand{\bl}{256}
\newcommand{\alp}{5e-7}
\newcommand{\bet}{1e-11}
\newcommand{\gam}{1e-18}
\newcommand{\ps}{1048576}     
\newcommand{\ns}{4194304}    

\begin{tikzpicture}[scale=0.7]
\begin{semilogxaxis}[
	     xlabel=Groups,
	     ylabel=Execution Time,
	     domain=1:\ps,
             log basis x=2,
	     legend style={at={(0.5,-0.2)},anchor=north,legend columns=2}
	]

\addplot[color=blue, mark=*] {2*(\ns^3)*\gam/\ps + (\ns/\bl) * ( ln(\ps)/ln(2) + 2*( sqrt(\ps)/sqrt(x) + sqrt(x) -2 ) )* \alp + 4*(\ns*\ns/sqrt(\ps))*(2 - sqrt(x)/sqrt(\ps) - 1/sqrt(x))*\bet };
\addplot[color=red, mark=x] {2*(\ns^3)*\gam/\ps+ (\ns/\bl) * ( ln(\ps)/ln(2) + 2*( sqrt(\ps) -1 ) )* \alp + 4*(\ns*\ns/sqrt(\ps))*(1 - 1/sqrt(\ps))*\bet };


\legend{HSUMMA, SUMMA}

\end{semilogxaxis}

\end{tikzpicture}
\caption{Prediction of SUMMA and HSUMMA on Exascale. Execution time in seconds. $p=1048576$}
\label{fig:theoretical_plot_exascale}
\end{figure}

Again we have $\frac{\alpha}{\beta}>2\frac{nb}{p}$ which means HSUMMA can be efficient and outperform SUMMA on exascale platforms and the theoretical plot 
is shown in Figure~\ref{fig:theoretical_plot_exascale}.

These analyses show that with any realistic platform parameters HSUMMA reduces the communication cost of SUMMA. However, one of the useful features of 
HSUMMA is that in the worst case it can use just one or $p$ group and have exactly the same performance as SUMMA.

\section{Conclusions}
We can conclude that our two-level hierarchical approach to parallel matrix multiplication significantly reduces
the communication cost on large platforms such as BlueGene/P.
Our experiments show that HSUMMA achieves $2.08$ times less communication time than SUMMA on $2048$ cores and $5.89$ times less communication cost on $16384$ cores.
Moreover, the overall execution time of HSUMMA is $1.2$ times less than the overall execution time of SUMMA on $2048$ cores,
and $2.36$ times less on $16384$ cores. This trend suggests that, while the number of processors increases our algorithm will be more scalable than SUMMA.
In addition, our experiments on Grid5000 show that our algorithm can be effective on small platforms as well.
All these results prove that whatever stand-alone application-oblivious optimized broadcast algorithms are made available for exascale platforms, 
they cannot replace application specific optimizations of communication cost.

At the moment, we select the optimal number of groups sampling over valid values. However, it can be easily automated and incorporated into the 
implementation by using few iterations of HSUMMA.

Our algorithm does not change the distribution of the matrices in SUMMA.
Currently, our algorithm  was designed for block-checkerboard distribution of the matrices. However,
we believe that by using block-cyclic distribution the communication can be better overlapped and parallelized
and thus the communication cost can be reduced even further. Thus, theoretical and practical analysis of our
algorithm with block-cyclic distribution is one of our main future works. In addition, until now we got all
these improvements without overlapping the communications on the virtual hierarchies.

We also plan to investigate the algorithm with more than two levels of hierarchy as we believe that in this case it is possible
to get even better performance. In addition, we plan to apply the same approach to other numerical linear algebra kernels such as QR/LU
factorization.


\section*{Acknowledgments}
The research in this paper was supported by IRCSET(Irish Research Council for Science,
Engineering and Technology) and IBM, grant numbers EPSG/2011/188 and
EPSPD/2011/207.

Some of the experiments presented in this paper were carried out using the Grid'5000 experimental testbed,
being developed under the INRIA ALADDIN development  action with support from CNRS, RENATER and several Universities as well
as other funding bodies (see https://www.grid5000.fr)

Another part of the experiments in this research were carried out using the resources of the Supercomputing Laboratory at King Abdullah University
of Science\&Technology (KAUST) in Thuwal, Saudi Arabia.




\bibliographystyle{IEEEtran}
\bibliography{IEEEabrv,icpp}

\begin{thebibliography}{10}
\providecommand{\url}[1]{#1}
\csname url@samestyle\endcsname
\providecommand{\newblock}{\relax}
\providecommand{\bibinfo}[2]{#2}
\providecommand{\BIBentrySTDinterwordspacing}{\spaceskip=0pt\relax}
\providecommand{\BIBentryALTinterwordstretchfactor}{4}
\providecommand{\BIBentryALTinterwordspacing}{\spaceskip=\fontdimen2\font plus
\BIBentryALTinterwordstretchfactor\fontdimen3\font minus
  \fontdimen4\font\relax}
\providecommand{\BIBforeignlanguage}[2]{{%
\expandafter\ifx\csname l@#1\endcsname\relax
\typeout{** WARNING: IEEEtran.bst: No hyphenation pattern has been}%
\typeout{** loaded for the language `#1'. Using the pattern for}%
\typeout{** the default language instead.}%
\else
\language=\csname l@#1\endcsname
\fi
#2}}
\providecommand{\BIBdecl}{\relax}
\BIBdecl

\bibitem{cannon:mmm}
L.~Cannon, ``A cellular computer to implement the kalman filter algorithm,''
  Ph.D. dissertation, Bozeman, MT, USA, 1969.

\bibitem{fox:mmm}
G.~C. Fox, S.~W. Otto, and A.~J.~G. Hey, ``Matrix algorithms on a hypercube i:
  Matrix multiplication,'' \emph{Parallel Computing}, vol.~4, pp. 17--31, April
  1987.

\bibitem{3d:mmm}
R.~Agarwal, S.~M. Balle, F.~G. Gustavson, M.~Joshi, and P.~Palkar, ``A
  three-dimensional approach to parallel matrix multiplication,'' \emph{IBM
  Journal of Research and Development}, vol.~39, 1995.

\bibitem{ChoiWD94:0}
J.~Choi, D.~W. Walker, and J.~Dongarra, ``Pumma: Parallel universal matrix
  multiplication algorithms on distributed memory concurrent computers,''
  \emph{Concurrency - Practice and Experience}, vol.~6, no.~7, pp. 543--570,
  1994.

\bibitem{summa:mmm}
R.~van~de Geijn and J.~Watts, ``Summa: Scalable universal matrix multiplication
  algorithm,'' \emph{Concurrency: Practice and Experience}, vol.~9, no.~4, pp.
  255--274, April 1997.

\bibitem{Dongarra:1997:SUG:265932}
L.~S. Blackford, J.~Choi, A.~Cleary, E.~D'Azeuedo, J.~Demmel, I.~Dhillon,
  S.~Hammarling, G.~Henry, A.~Petitet, K.~Stanley, D.~Walker, and R.~C. Whaley,
  \emph{ScaLAPACK user's guide}, J.~J. Dongarra, Ed.\hskip 1em plus 0.5em minus
  0.4em\relax Philadelphia, PA, USA: Society for Industrial and Applied
  Mathematics, 1997.

\bibitem{Solomonik:2011:CPM:2033408.2033420}
\BIBentryALTinterwordspacing
E.~Solomonik and J.~Demmel, ``Communication-optimal parallel 2.5d matrix
  multiplication and lu factorization algorithms,'' in \emph{Proceedings of the
  17th international conference on Parallel processing - Volume Part II}, ser.
  Euro-Par'11.\hskip 1em plus 0.5em minus 0.4em\relax Berlin, Heidelberg:
  Springer-Verlag, 2011, pp. 90--109. [Online]. Available:
  \url{http://dl.acm.org/citation.cfm?id=2033408.2033420}
\BIBentrySTDinterwordspacing

\bibitem{exascale:report}
M.~Kondo, ``Report on exascale architecture roadmap in {J}apan,'' 2012.

\bibitem{Beaumont01matrixmultiplication}
O.~Beaumont, V.~Boudet, F.~Rastello, and Y.~Robert, ``Matrix multiplication on
  heterogeneous platforms,'' 2001.

\bibitem{alexey:hphc}
A.~Lastovetsky and J.~Dongarra, \emph{High Performance Heterogeneous
  Computing}.\hskip 1em plus 0.5em minus 0.4em\relax Wiley, 2009.

\bibitem{Snir:1998:MCR:552013}
M.~Snir, S.~Otto, S.~Huss-Lederman, D.~Walker, and J.~Dongarra, \emph{MPI-The
  Complete Reference, Volume 1: The MPI Core}, 2nd~ed.\hskip 1em plus 0.5em
  minus 0.4em\relax Cambridge, MA, USA: MIT Press, 1998.

\bibitem{Thakur03improvingthe}
R.~Thakur, ``Improving the performance of collective operations in mpich,'' in
  \emph{Recent Advances in Parallel Virtual Machine and Message Passing
  Interface. Number 2840 in LNCS, Springer Verlag (2003) 257–267 10th
  European PVM/MPI User’s Group Meeting}.\hskip 1em plus 0.5em minus
  0.4em\relax Springer Verlag, 2003, pp. 257--267.

\bibitem{Barnett94interprocessorcollective}
M.~Barnett, S.~Gupta, D.~G. Payne, L.~Shuler, R.~Geijn, and J.~Watts,
  ``Interprocessor collective communication library (intercom),'' in \emph{In
  Proceedings of the Scalable High Performance Computing Conference}.\hskip 1em
  plus 0.5em minus 0.4em\relax IEEE Computer Society Press, 1994, pp. 357--364.

\bibitem{Thakur05optimizationof}
R.~Thakur and R.~Rabenseifner, ``Optimization of collective communication
  operations in mpich,'' \emph{International Journal of High Performance
  Computing Applications}, vol.~19, pp. 49--66, 2005.

\bibitem{633174}
D.~Scott, ``Efficient all-to-all communication patterns in hypercube and mesh
  topologies,'' in \emph{Distributed Memory Computing Conference, 1991.
  Proceedings., The Sixth}, apr-1 may 1991, pp. 398 --403.

\bibitem{Hockney:1994:CCM:179977.179986}
\BIBentryALTinterwordspacing
R.~W. Hockney, ``The communication challenge for mpp: Intel paragon and meiko
  cs-2,'' \emph{Parallel Comput.}, vol.~20, no.~3, pp. 389--398, Mar. 1994.
  [Online]. Available: \url{http://dx.doi.org/10.1016/S0167-8191(06)80021-9}
\BIBentrySTDinterwordspacing

\bibitem{Shroff00collmark:mpi}
M.~Shroff and R.~A. V.~D. Geijn, ``Collmark: Mpi collective communication
  benchmark,'' Tech. Rep., 2000.

\bibitem{Grebovic:2007}
J.~Pje\u{s}ivac-Grbovi\'{c}, ``Towards automatic and adaptive optimizations of
  {M}{P}{I} collective operations,'' Ph.D. dissertation, University of
  Tennessee, Knoxville, December, 2007.

\bibitem{Patarasuk:2008:TPB:1370314.1370586}
\BIBentryALTinterwordspacing
P.~Patarasuk, X.~Yuan, and A.~Faraj, ``Techniques for pipelined broadcast on
  ethernet switched clusters,'' \emph{J. Parallel Distrib. Comput.}, vol.~68,
  no.~6, pp. 809--824, Jun. 2008. [Online]. Available:
  \url{http://dx.doi.org/10.1016/j.jpdc.2007.11.003}
\BIBentrySTDinterwordspacing

\bibitem{Balaji:mpibg}
\BIBentryALTinterwordspacing
P.~Balaji, R.~Gupta, A.~Vishnu, and P.~Beckman, ``Mapping communication layouts
  to network hardware characteristics on massive-scale blue gene systems,''
  \emph{Comput. Sci.}, vol.~26, no. 3-4, pp. 247--256, Jun. 2011. [Online].
  Available: \url{http://dx.doi.org/10.1007/s00450-011-0168-y}
\BIBentrySTDinterwordspacing

\end{thebibliography}
%




\end{document}